\documentstyle[prl,aps,twocolumn,epsf]{revtex}

\include{psfig}

\title{
Resonant transmission of normal electrons through Andreev states in 
ferromagnets}  
\author{A. Kadigrobov$^{1,2}$, R. I. Shekhter$^{1}$, M.~Jonson$^{1}$,\\
Z. Ivanov$^{3}$, and T. Claeson$^{3}$ }
\address{
 $^{1}$Department of Applied Physics,
 Chalmers University of Technology and G\"oteborg University, 
SE-412 96 G\"oteborg, Sweden\\
 $^{2}$B. I. Verkin Institute for Low Temperature Physics \& 
Engineering, \\
National Academy of Science of Ukraine, 47 Lenin Ave., 310164 Kharkov, 
Ukraine\\
$^{3}$Department of Physics, Chalmers University of Technology, 
S-412 96 G\"oteborg, Sweden}

\begin{document}

\date{January 20, 1999}
\maketitle

\begin{abstract}
Giant oscillations of the conductance of a superconductor - ferromagnet - 
superconductor Andreev interferometer are predicted. The effect is due to the 
resonant transmission of normal electrons through Andreev levels  
when the voltage $V$ applied to the ferromagnet is close to $2h_0/e$ ($h_0$ is 
the spin-dependant part of the electron energy). The effect of bias voltage
and phase difference between the superconductors on the 
current and the differential conductance is presented. 
These efects allow a direct spectroscopy of Andreev levels in the ferromagnet.
\end{abstract} 

Recently a high sensitivity of the conductance of mesoscopic systems to the 
superconductor phase difference has been observed and theoretically 
considered in superconductor - normal
conductor - superconductor heterostructures (S/N/S heterostructures) 
(see, e.g., the review paper by Lambert and Raimondi \cite{Lamb}). This effect 
arises because 
of a  quantum interference of quasiparticles due to Andreev scattering
 at two (or more) N-S interfaces. This is caused by the fact that the phase 
of the superconducting condensate is imposed on the quasiparticle wave 
function in the normal metal. One of the manifestations of the quantum 
interference is giant oscillations of the conductance of the normal metal 
as a function of the phase difference between the superconductor predicted 
in \cite{GO,GOdif}. 

A single electron in a normal metal  with energy below the superconductor 
energy gap, $\Delta$,
can not penetrate into the superconductor. However, under Andreev reflection 
at an N-S interface 
two electrons with nearly opposite momenta and  spins  leave the normal 
metal to create a Cooper pair in the superconductor; hence the incident 
electron is transformed into a hole with the opposite direction of the  spin. 
The spin flip does not effect the interference pattern 
of the non-magnetic normal metal because all  energy levels are doubly 
degenerate with respect to spin. In ferromagnets, however, 
this degeneracy is lifted due to the interaction of the electron spin with 
the ferromagnet's spontaneous moment (below we refer to it as the 
exchange-interaction energy $h_0$), and   
electrons with opposite spins occupy different energy 
bands (Fig. 1).  
In this case,  the change of spin direction associated with  Andreev 
scattering shifts the reflected quasiparticle from one band to the other. 
The latter   influences the quantum interference. The Josephson current in a 
superconductor-ferromagnet-superconductor (S/F/S)
structure was  investigated in 
Refs.~\cite{Bulaevskii,Falko,Demler}; transport properties of F/S junctions 
were investigated in 
\cite{Beenakker2,Dong,Shi,Vasko,Giroud,Shashi,Petrashov3,Falko1}; 
experiments on the boundary resistance of an S/F/S system was reported 
in \cite{Fierz}, and phase coherent effects in the conductance of a 
ferromagnet contacting a superconductor  were observed in \cite{Petrashov}.
\begin{figure}
\centerline{\psfig{figure=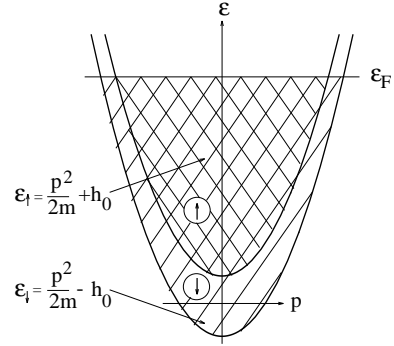,width=5cm}}
\vspace*{2mm}
\caption{Energy bands for electrons with opposite spins
}
\end{figure}
In this Letter we predict  giant oscillations  in the conductance of an
S/F/S heterostructure, of 
the Andreev interferometer type, in which the ferromagnet part is separated 
from the reservoirs of normal electrons with potential barriers ("beam 
splitters") of low transparency, $t_r \ll 1$, (see the insert in Fig. 2).

In the case of  Andreev reflections, the paramagnetic effect  essentially 
modifies the interference pattern in the ferromagneict region. The momentum of 
an electron with spin up/down),  $p^{e}_{\uparrow}$ /$p^{e}_{\downarrow}$, 
and the momentum of the reflected hole with the spin down/up, 
$p^{h}_{\uparrow}$/   $p^{h}_{\downarrow}$  are  (see \cite{Beenakker2}).
\begin{equation}
p^{(e)}_{\uparrow \downarrow}=\sqrt{p_F^2 +2m(E \pm h_0});\;\; 
p^{(h)}_{\downarrow \uparrow}=\sqrt{p_F^2 -2m(E\pm h_0})
\label{magmomenta}
\end{equation}
where $E$   is the energy of the incident electron measured from the Fermi 
level $\epsilon_F$, $p_F$ is the Fermi momentum, and $m$ is the electron mass.

From Eq. (\ref{magmomenta}) it follows that in contrast to the  non-magnetic 
case,
near the Fermi level  ($E \approx 0$) the electron and the hole momenta in 
the ferromagnet are  different, and for large enough $h_0$ (usually $h_0$ is 
greater than the Thouless energy) the interference effects are absent due to 
the destructive interference.
This fact demonstrates the conflict between superconductivity and magnetic 
ordering in S/F/S structures.
\begin{figure}
\centerline{\psfig{figure=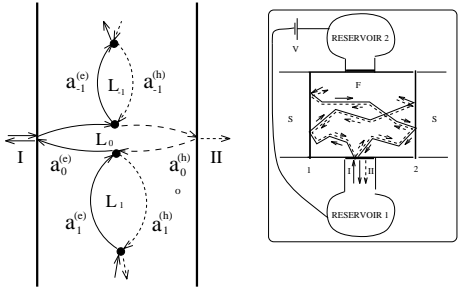,width=6cm}}
\vspace*{2mm}
\caption{Schematic representation of resonant transmission of an incident 
electron
which  tunnels through  potential barrier I, moves along a 1D disordered 
chain of  scatterers (dots) where Andreev  scattering (back-scattering) + 
normal  scattering 
(forward scattering) takes place, and is reflected back through the first 
barrier I as an electron
and transmitted through the second barrier II as a hole. 
Semi-classical electron  and hole paths  are  shown with full and dashed
lines, respectively.   Trajectory sections connecting successive scattering 
events at 
different N/S-interfaces have lengths  $ L_i$;  $i=0, \pm1,...$. The 
insertion schematically shows the geometry of the system under consideration 
and a classical 
trajectory contributing to the resonant part of the conductance.  Thick lines 
 indicate potential barriers.}
\end{figure}
However, interference effects in the ferromagnet can exist albeit  at some 
finite voltage $V$ applied    between the reservoirs. If the energy $|E| 
\approx h_0<|\Delta|$  the change of  the quasi-particle momentum under 
Andreev reflection is small (see Eq. (\ref{magmomenta})), while the velocity 
changes its sign, and an essential  cancellation of the phase gain along 
trajectories including electron-hole transformations at the superconducting 
boundaries takes place.  At  $E= h_0$  any such a classical trajectory is 
closed (in this case, under Andreev reflection  the electron and hole  
momenta are equal and hence the reflected quasi-article is sent exactly back
   along the classical path of the incident quasiparticle), and  
 this cancellation is complete  at $\phi=\pi(2l+1), \;l= 0, \pm 1, ...$ 
irrespective of the geometry and the length of the trajectory 
\cite{magfield}. From here it follows that at $|E-h_0| \ll E_{Th}$  and 
$\phi$ close to  odd numbers of $\pi$,  such paths take part in the 
constructive interference resulting in resonant transmission through Andreev 
levels. 
In our calculations of the probability amplitude of the electron-hole 
reflection back to the reservoir of the electron injection \cite{eltramsm} 
we use the  approach developed by us in Ref.~\cite{Bagwell} assuming the
motion of quasi-particles inside the ferromagnet to be semi-classical  
(this assumption is valid if the de-Broglie wave length $\lambda_F$ of 
electrons is the shortest length in the ferromagnet).  Within  this approach 
one can find the wave function of the scattered quasiparticles by mapping 
the incident wave along  classical  
paths determining the phase of rapid oscillations $\Theta =S/\hbar$ as a 
classical action $S = \int pdl$ along the path.
A typical classical trajectory of this kind for an incident electron which 
undergoes a number of Andreev
and normal reflections at F-S boundaries  is shown in the insert of Fig. 2
(solid and dashed lines are for electronic and hole paths, respectively).
The electron-hole transmission along this trajectory is similar to the
resonant transmission of an electron through a two-barriers system 
(schematically shown in Fig. 2) in which the incident electron  tunnels  
through a potential barrier I (solid line I), moves along  
a one-dimensional chain of scatterers (black dots in Fig. 2 representing
Andreev and normal reflections at F/S interfaces), and then is reflected 
back as an electron
through potential barrier I and  transmitted 
through potential barrier II as a hole.  $L_i$ ($i=0,\pm 1, ...$) is the 
length of the quasi-particle path between two successive scatterings at F/S 
interfaces
which is the distance between the neighboring scatterers for the 1D chain of 
Fig.~2. The paths $L_i$  are uncorrelated and hence the chain of Fig.2 is
a 1D system with disordered distances between the scatterers. In the same 
way  as in  Ref.~\cite{Bagwell}, it can be shown
 that due to the above-mentioned phase compensation 
the motion of the quasi-particle in this chain is reduced to the conventional 
quantum motion of  an electron with energy $E-h_0$ (but having the Fermi  
velocity $\sim v_F$) in the 1D disordered chain of centers of back-scatterings
where the back-scattering  amplitude is the probability amplitude of the 
Andreev reflection
$r_A^{(1,2)}$ and the amplitude to pass to the next section of the chain is 
the probability amplitude of the normal reflection $r_N^{(1,2)}$ at F/S 
interfaces 1 and 2 (the probability amplitudes $r_A$ and $r_N$ are given in 
\cite{Beenakker2,Shelankov,Blonder}).
In this situation, for $E\neq h_0$ the phase gains between successive 
back-scatterings are random, and quasi-particle
localization 
takes place.  For $|r_N^{(1,2)}|\ll 1$ \cite{normref} and  
$t_r \ll 1$ a sharp resonant transmission between points   I and II through  
discrete energy levels 
(of the Andreev type) which correspond to the quasiparticle states localized
around the section of the electron injection, occurs.  Matching amplitudes 
$a_i^{e,h}$ at the centers of scattering (dots in Fig. 2) and taking into 
account the phase gains
along the paths between them 
show  the probability of 
electron-hole resonant transmission through an energy level $E_{\alpha}$ 
\cite{transp} to be of  the Breit-Wigner form,
$
T(E,\alpha)\propto t_r^2/
[\left((E-E_{\alpha})\tau_0\right)^2/\hbar^2 + t_r^2 \times{\rm const.}],
$
 where  $\tau_0$ is the time of motion in 
the section of injection. 
 
The total electron-hole transmission probability $T_{eh}(E)$ is a sum of
$T(E,\alpha)$ with respect to the starting points of the
semi-classical trajectories inside the reservoir separated by the distance 
of the order of $\lambda_F$. These trajectories  meet different "random" sets
of impurities, and hence  their path lengths 
and  the times of quasiparticle   propagation  along them
 are randomly distributed.  Therefore, the summation over the starting points 
is equivalent to averaging
the transmission probability with respect to realizations of times  $\tau_i$  
($\tau_i$ is the time of propagation along section $i$) 
(see \cite{GOdif,Bagwell}). It seems reasonable to assume the 
propagation times $\tau_i$ to be uncorrelated. Under this 
assumption, as is shown in \cite{Bagwell}, the total transmission  probability
 $T_{eh}(E)$  is proportional to   the
density of localized states in  the 1D disordered chain of Fig. 2, and 
using   the Lambert
formula \cite{Lambert} one gets  the  transport current for temperature  
$T=0$  as 
\begin{equation}
I =\left(t_r N_{\perp}e/h \right)E_{Th}\sum_{\uparrow, 
\downarrow}\int_{-eV/2}^{eV/2}  <\nu_{rand}^{(\uparrow, \downarrow)}(E)>dE
\label{current}
\end{equation}
(here and below we assume 
$t_r\ll (|r_N^{(1)}|+ |r_N^{(2)}|)/2\ll 1$). In Eq. (\ref{current}) 
$N_{\perp}=S/\lambda_F^2$, $S$ is the F/S contact area, $\lambda_F$ is the 
electron wave length, $<\nu_{\uparrow, \downarrow}^{rand}(E)>$ is the  
density of states for a quasi-particle with the spin up ($\uparrow$) or 
down ($\downarrow$) averaged with respect to the configurations of  ${\tau}_n$.

Now  we assume  the distribution $P(\tau)$ for the propagation  times to be 
of the
Lorentzian form $$P(\tau)=\gamma/\pi[(\tau-\bar{\tau})^2 + \gamma^2]$$
($\bar{\tau }= L_S^2/D$ and $L_S$ is the distance between the superconductors)
that,  for the configuration of Fig. 2,  permits to find the density of state
exactly.
Using Eq. (\ref{current}) one 
finds the resonant phase-sensitive part of  the differential conductance of 
the system $G= d I/dV$ to be
\begin{equation}
\begin{array}{l}
G=\frac{\sqrt{2}e^2}{h}N_{\perp}t_r |\bar{V}| \times\\
\left\{\frac{\sqrt{[4\bar{V}^4+\epsilon_{a}^4]
[4\bar{V}^4+\epsilon_{b}^4]}+
\epsilon_{a}^2 \epsilon_{b}^2-4\bar{V}^4)}
{[4\bar{V}^4+\epsilon_{a}^4][4\bar{V}^4+\epsilon_{b}^4]} \right\}^{1/2}
\end{array}
\label{difcond}
\end{equation}
where   $\bar{V} = (eV/2 -h_0)/E_{Th}$,
$\epsilon_{a,b}=[\delta\phi^2+(|r_N^{(1)}|\pm|r_N^{(2)}|)^2]^{1/2}$ is the 
dimensionless applied voltage measured from $h_0$, 
$\delta\phi$ is the minimal value of $|\phi -\pi (2l+1)| $, $l =0, \pm1,
\pm2,...$.
While writing Eq. (\ref{difcond}) we took  $\gamma = \bar{\tau}$, and  
assumed  $|h_0-eV/2|\ll E_{Th}$. Eqs.~(\ref{current}) and (\ref{difcond}) 
describe the the current and differential conductance at $T=0$ for both the 
magnetic $h_0 \neq 0$ and non-magnetic 
$h_0 = 0$ cases.

Numerical results  for the conductance and the current  based on 
Eq. (\ref{difcond}) are shown in Figs.~3 and 4.
They demonstrate a high sensitivity of the conductance and the non-linear
current-voltage characteristics
to both the superconductor phase difference $\phi$ and the applied voltage 
$V$. 

At odd multiples of $\pi$ and $|r_N^{(1)}| = |r_N^{(2)}|$  there is a symmetry 
between the clock- and counter-clockwise motions of electron-hole pairs in 
the ferromagnet, and   the energy level $E= h_0$ is degenerate (see above) 
\cite{Spivak}.  Under this condition   the maximum of the resonant 
transmission through Andreev levels  is at $ eV/2 = h_0$, and a resonant 
peak in the conductance is observed   (Fig. 3a). Even a small deviation of 
$\phi$ from an odd multiple of
 $\pi$,  will repel Andreev levels from $h_0$, and 
the conductance peak splits in two peaks. 

 In a more realistic experimental case when   $|r_N^{(1)}| \neq |r_N^{(2)}|$
the symmetry is broken, and Andreev levels are repelled from the level $h_0$
(see \cite{Beenakker3,GOdif}),   the shift being proportional to 
$\delta r_N =||r_N^{(1)}| -|r_N^{(2)}||$.  As a result the resonant peaks of 
the conductance are split
as shown in Fig. 3b. 
\begin{figure}
\centerline{\psfig{figure=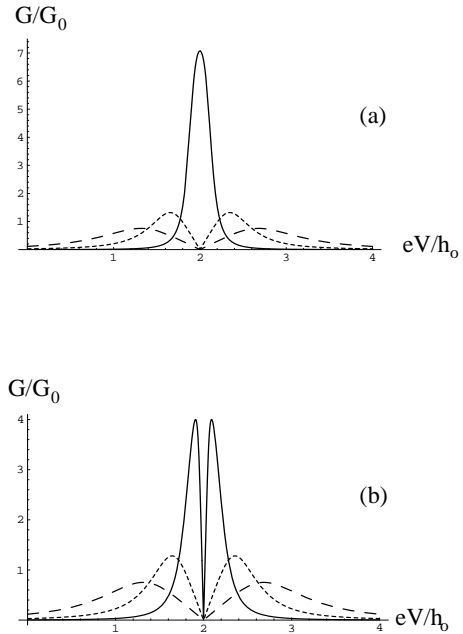,width=6cm}}
\vspace*{2mm}
\caption{Normalized differential conductance $G=dI/dV$ of the S/F/S structure 
for $|r_N^{(1)}|=|r_N^{(2)}|=0.1$ and $|r_N^{(1)}|=0.05,\;|r_N^{(2)}|=0.1$ 
shown in Fig. 3a and Fig3b, respectively, at  phase differences $\phi=\pi$ 
(full line),  $\phi=1.1\pi$ (dotted line) and  $\phi=1.2\pi$ (dashed line); 
$G_0=(\sqrt{2} e^2/h)N_{\perp}t_r$.
}
\end{figure}

Current-voltage characteristics for different $\phi$  are presented in Fig. 4.
\begin{figure}
\centerline{\psfig{figure=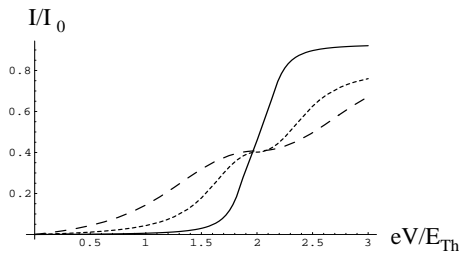,width=6cm}}
\vspace*{2mm}
\caption{Normalized current-voltage characteristics for phase differences 
$\phi=\pi$ (full line),  $\phi=1.1\pi$ (dotted line) and  $\phi=1.2\pi$ 
(dashed line) shown for $|r_N^{(1)}|=0.05,\;|r_N^{(2)}|=0.1$ and 
$h_0=E_{Th}$; $I_0=(\sqrt{2}e^2/2\pi\hbar)N_{\perp}t_r(2h_0/e)$. 
}
\end{figure}
At low voltages, far from $2h_0/e$, we have a resonant tunneling of 
quasi-particles through separate Andreev levels, and the current level is 
low. When $eV/2 \approx h_0$ and $\phi = \pi$ Andreev levels concentrate 
near $h_0$, and we have simultaneous resonant transport through the 
whole number of  $N_{\perp}$ states resulting in a jump of the current 
$\Delta I = ||r_N^{(1)}| + |r_N^{(2)}|| G_{max} h_0/2e$ ($G_{max}$ is the 
maximal value of the conductance).
 When 
$\phi$ deviates from $\pi$ the number of Andreev levels concentrated near 
$h_{0}$ is decreasing that results in a decrease of the sensitivity of the 
current to the voltage.

We note here that the  curve for the differential conductance $G$ as a 
function of $eV$ repeats the density of Andreev states in the diffusive 
ferromagnet permitting a direct spectroscopy of the Andreev levels by 
conductance and current measurements.

In conclusion we have demonstrated a pronounced  possibility for spectroscopy 
of Andreev states in ferromagnets at energies even greater than the Thouless 
energy.  The paramagnetic effect determines sharp peaks
for the conductance as a function of the superconductor phase difference 
$\phi$ and the applied voltage $V$ near $\phi =\pi(2l+1),\;l=0,\pm1,\pm2, ...$ 
and $V=2h_0 /e$, respectively. This phenomenon is a convenient tool for the 
Andreev level spectroscopy, and enables applications, e.g. as a double-gate 
ferromagnet transistor and an AND logic element as described in 
\cite{patent}.   
   
This work was supported by  the Swedish KVA and NFR and by the National
Science Foundation under Grant No. PHY94-07194.


\begin{thebibliography}{99}

\bibitem{Lamb} C. J. Lambert and R. Raimondi,  J. Phys.: Condens. Matter, 
{\bf 10}, 901 (1997).

\bibitem{GO} A. Kadigrobov, A. Zagoskin, R. I. Shekhter, M. Jonson, Phys. 
Rev. B {\bf52}  R8662 (1995).

\bibitem{GOdif}  H. A. Blom, A. Kadigrobov, A. Zagoskin, R. I. Shekhter, and
M. Jonson, 
Phys. Rev. B  {\bf 57}, 9995 (1998).

\bibitem{Bulaevskii} L. N. Bulaevski, A. I. Buzdin and S. V. Panjukov, 
Solid State Commun. {\bf 44}, 539 (1982).

\bibitem{Falko} S. V. Kuplevakhskii and I. I. Falko, Pis'ma Zh. Eksp. Teor. 
Fiz. {\bf 52}, 957, (1990) (JETP Lett. {\bf 52}, 340 (1990)).

\bibitem{Demler} E. A. Demler, G. B. Arnold, M. R. Beastly, Phys. Rev. 
B{\bf 55}, 15174 (1997).

\bibitem{Beenakker2}  M. J. M. de Jong and C. W. J. Beenakker, Phys. Rev. 
Lett. {\bf 74}, 1657 (1995).

\bibitem{Dong} Z. W. Dong et al., Appl. Phys. Lett. {\bf 71}, 1718 (1997). 



\bibitem{Shi} Junren Shi, Jinming Dong, D. Y. Xing, Physica C {\bf 282-287},
1853 (1997).

\bibitem{Vasko} V. A. Vasko {\it et al}., Appl. Phys. Lett., {\bf 73}, 844 
(1998).

\bibitem{Giroud} M. Giroid {\it et al.}, Phys. Rev. B, {\bf 58}, R11 872 
(1998).

\bibitem{Shashi} Shashi K. Upadhyay {\it et al.}, Phys. Rev. Lett. {\bf 81}, 
32487 (1998)

\bibitem{Petrashov3} V. T. Petrashov {\it et al.}, The Vth Bristol workshop on
the Bogoliubov - de Gennes Equations", Bristol, November 20-22 (1998). 

\bibitem{Falko1}  Vladimir I. Falko, A. F. Volkov, and C. Lambert, 
cond-mat/9901051, 7 January 1999.

\bibitem{Fierz} C. Fierz, S.-F. Lee, J. Bass, W. P. Pratt, Jr., and P. A. 
Schroeder, J. Phys. Condens. Matter {\bf 2}, 9701 (1990).


\bibitem{Petrashov} V. T. Petrashov, V. N. Antonov, S. V. Maksimov, and 
R. Sh. Shakhadarov, Pis'ma Zh. Eksp. Teor. Fiz. {\bf 59}, 523 (1994) 
(JETP Lett. {\bf 59} 551 (1994)).

\bibitem{magfield}  Here and below we neglected the effect of the intrinsic 
magnetic field of the ferromagnet on  the  electron and hole motion. 
Reversibility of the electron - hole trajectories is violated in magnetic 
fields
$H \gg H_c$ where $H_c = \Phi_0 l_i^2/L^4$ ($\Phi_0$ is the flux quantum, 
$l_i$ is the electron free path length, and $L$ is the distance between 
the N-S interfaces).
Estimations show that for  parameter experimentally available $H_c$ may be 
of several tens  Oe, while for a ballistic ferromagnet $H_c$ may be of 
several kOe.



\bibitem{eltramsm} It can be shown that for the case of the resonant 
transmission under consideration the contribution of the electron-electron 
transmission to the conductance is exactly the same as from the electron-hole 
reflection, and hence below we consider only the latter one.

\bibitem{Bagwell} A. Kadigrobov, L. Y. Gorelik, R. I. Shekhter, M. Jonson,
cond-mat/9811212 (16 November 1998).


\bibitem{Shelankov} A. L. Shelankov, Pis'ma Zh. Eksp. Teor. Fiz. {\bf 32}, 
122 (1980) (JETP Lett.)

\bibitem{Blonder} G. E. Blonder, M. Tinkham, and T. M. Klapwijk, Phys. Rev. 
B{\bf 25} 4515 (1982).


\bibitem{normref} As was pointed out by Jong and Beenakker 
\cite{Beenakker2} the
jump of the magnetization at the ferromagnet-superconductor boundary 
results in the back scattering the probability of which is of the order of 
$(h_0/\epsilon_F)^2$.

\bibitem{transp} In our case  the transparencies of the barriers are low  
and the dispersion in the distribution of times of motion inside the wells 
is of the order of the mean value $\bar{\tau}$. Hence, the localization 
radius is of the order of  $\bar{{\cal L}}=v_F\bar{\tau}$, and for a given 
time configuration
there is only one resonant level in the energy range of interest $|E -h_0| 
\sim |r_N|E_{Th}\ll E_{Th}$.

\bibitem{Lambert} C. J. Lambert, J. Phys.: Cond. Matter {\bf 3}, 6579 (1991);
Cond. Matter {\bf 5}, 707 (1993).

\bibitem{Spivak} In the absence of magnetic ordering this degeneration takes 
place at the Fermi level \cite{GOdif} that   agrees with paper \cite{Spivak1} 
where a peak
in the density of  Andreev states was found at $\phi = \pi$.

\bibitem{Spivak1}  F. Zhou, P. Charlat, B. Spivak, B. Pannetier, Journal of 
Low Temperature Physics, {\bf 110}, 841 (1998).

\bibitem{Beenakker3} C. W. J. Beenakker, Phys. Rev. Lett. {\bf 67}, 602 (1991).

\bibitem{patent} Z. Ivanov, R. Shekhter, A. Kadigrobov, T. Claeson, M. Jonson,
E. Wikborg, Patent "Superconducting transistor", N 9804088 - 4, 
registered 27 November 1998.


\end{thebibliography}
\end{document}